\documentclass{ifacconf}
\usepackage{amssymb}
\usepackage{graphicx}      
\usepackage{natbib}        
\usepackage{xcolor}
\usepackage{soul}
\usepackage{amsmath}

\begin{document}
\begin{frontmatter}

\title{LLM-Enhanced Symbolic Control for Safety-Critical Applications} 


\author[First]{Amir Bayat} 
\author[Third]{Alessandro Abate}
\author[Second]{Necmiye Ozay}
\author[First]{Raphaël M. Jungers.}

\address[First]{ICTEAM Institute, UCLouvain,
	Louvain-la-Neuve, Belgium (e-mail: FirstName.LastName@ uclouvain.be)}
\address[Third]{University of Oxford, 
	Oxford, UK, (e-mail: Alessandro.Abate@cs.ox.ac.uk)}
\address[Second]{University of Michigan, 
   Ann Arbor, Michigan, USA (e-mail: necmiye@umich.edu)}

\begin{abstract}                
Motivated by Smart Manufacturing and Industry 4.0, we introduce a framework for synthesizing Abstraction-Based Controller Design (ABCD) for reach-avoid problems from Natural Language (NL) specifications using Large Language Models (LLMs). A Code Agent interprets an NL description of the control problem and translates it into a formal language interpretable by state-of-the-art symbolic control software, while a Checker Agent verifies the correctness of the generated code and enhances safety by identifying specification mismatches. Evaluations show that the system handles linguistic variability and improves robustness over direct planning with LLMs. The proposed approach lowers the barrier to formal control synthesis by enabling intuitive, NL-based task definition while maintaining safety guarantees through automated validation.
\end{abstract}


\end{frontmatter}

\section{Introduction}
Many real-world control applications, such as autonomous vehicles, medical devices, and industrial process control systems, are safety-critical and demand stringent guarantees. In such systems, violations of operational constraints can lead to catastrophic or irreversible outcomes. To ensure both safety and performance, control strategies must be designed with formal guarantees. However, the complexity of cyber-physical systems (CPSs)—which often include nonlinear, hybrid, or high-dimensional dynamics—makes control synthesis a challenging task.

A promising solution is Abstraction-Based Controller Design (ABCD), also known as symbolic control \cite{tabuada2009verification, zamani2014symbolic}. In this approach, the original dynamical system is approximated by a finite, discrete abstraction. A controller is synthesized at this abstract level and then concretized to operate on the original system while preserving correctness. ABCD provides correct-by-design guarantees and has shown success in handling complex specifications, such as those expressed in temporal logic. However, implementing these methods is non-trivial: it typically requires deep domain expertise, significant effort, and careful tuning. Moreover, scalability remains a fundamental challenge due to the well-known curse of dimensionality.

In classical symbolic control, it is the engineer’s task to interpret diverse constraints, such as physical limitations, budget requirements, and sustainability goals, and convert them into a set of formal specifications understandable by the machine. This translation process can be extremely time-consuming, or even infeasible, especially for non-expert users or in systems with evolving requirements. Among various safety-critical applications, smart manufacturing and Industry 4.0 highlight the need for intelligent, flexible interfaces that accommodate decentralized operation and operators with varying expertise and terminology.

Nevertheless, recent advances in optimization, learning, and data-driven methods have triggered a revival of ABCD approaches \cite{badings2023robust, nilsson2017augmented, zamani2014symbolic, haesaert2017verification, banse2023data, calbert2024smart}. In particular, \texttt{Dionysos}\footnote{\texttt{https://github.com/dionysos-dev/Dionysos.jl}} \cite{calbert2024dionysos}, a modular and open-source software tool, has been developed to facilitate control synthesis using symbolic methods. \texttt{Dionysos} integrates modern optimization techniques with smarter abstraction strategies, making it feasible to address more complex control problems that were previously out of reach.

At the same time, Large Language Models (LLMs)—a class of machine learning models developed for natural language processing—have surged in popularity due to their intuitive interfaces and impressive capabilities in intent detection \cite{arora2024intent, Njah2025}, code generation \cite{jiang2024survey}, and general knowledge synthesis. Despite these strengths, LLMs face significant challenges when it comes to reasoning about physical and dynamical systems, largely because they are trained solely on text and lack grounding in real-world dynamics \cite{ahn2022can}. As a result, their direct application to engineering problem-solving remains limited. However, performance improves markedly when complex tasks are decomposed into simpler subtasks \cite{khot2022decomposed}. This insight has inspired frameworks like ControlAgent \cite{guo2024controlagent}, which coordinates multiple few-shot-trained LLM agents to synthesize PID controllers from natural language (NL) specifications. While promising, such methods are currently restricted to linear systems and classical control.

Other lines of work have used LLMs to translate NL into formal specifications such as temporal logic \cite{chen2024autotamp, pan2023data}, but, to the best of our knowledge, none have explored direct integration with ABCD pipelines or aimed to support full controller synthesis for nonlinear or hybrid systems.

In this work, we take a first step toward automated, end-to-end abstraction-based controller synthesis guided by NL. Specifically, we leverage LLMs as few-shot learners to extract the specifications of reach-avoid problems from NL descriptions and automatically generate code based on these specifications. The generated code fills a modular component within a larger framework that solves reach-avoid problems symbolically. To evaluate the effectiveness and robustness of our method, we introduce a benchmark dataset composed of diverse environments with varying obstacle configurations and target regions. This work can also be seen as an early step toward meeting emerging needs in smart manufacturing by exploring how LLMs can help interpret NL specifications and support the synthesis of safe planners/controllers.

\section{Preliminary}

This section provides the necessary background by briefly introducing the \texttt{Dionysos} toolbox and describing the classical abstraction method it implements for controller synthesis.

\subsection{Dionysos}
 
As previously mentioned, \texttt{Dionysos} is a state-of-the-art modular package designed to solve optimal control problems for complex dynamical systems using ABCD methods. Given the system dynamics and problem specifications, it can automatically construct the abstraction, synthesize an abstract controller, and concretize it to operate on the concrete system. In a reach-avoid problem, these spescifications involve the state space bounds, initial state, target states and constraints. 

\texttt{Dionysos} supports multiple abstraction techniques, ranging from classical methods to more advanced approaches such as ellipsoidal abstractions, hierarchical abstractions, lazy abstractions, and lazy ellipsoidal abstractions. In this work, the focus is placed on solving problems using the \emph{Uniform Grid Abstraction} method. A concise overview of this method is provided in the following subsection.

\subsection{Classical Abstraction-Based Controller Design}

In this section, we provide a concise overview of classical ABCD, which serves as the foundation for the control synthesis techniques implemented in \texttt{Dionysos}. The ABCD methodology proceeds through the following main stages:

\subsubsection{Defining the System Model}

We consider a continuous-time control system described by:
\begin{equation} \label{eq:1}
	\dot{x}(t) = f(x(t), u(t)),
\end{equation}
where \( x(t) \in \mathcal{X} \subseteq \mathbb{R}^n \) is the system state and \( u(t) \in \mathcal{U} \subseteq \mathbb{R}^m \) is the control input. The function \( f: \mathcal{X} \times \mathcal{U} \rightarrow \mathbb{R}^n \) defines the system dynamics.

A \emph{specification} \( \Sigma \subseteq \mathcal{X}^{\mathbb{R}_{\geq 0}} \) defines a set of admissible trajectories. The control problem is to design a feedback controller $u:\mathcal{X} \rightarrow \mathcal{U}$ such that the closed-loop trajectories satisfy $\Sigma$.

In this work, we focus on \emph{reach-avoid specifications}, of the following form:
\begin{multline}\label{eq:2}
	\Sigma^{\text{reach}} = \big\{ x(\cdot) \in \mathcal{X}^{\mathbb{R}_{\geq 0}} \,\big|\,
	x(0) \in \mathcal{X}_{\text{in}} \Rightarrow \\
	\big( \exists t_f \in \mathbb{R}_{\geq 0}: x(t_f) \in \mathcal{T} \big) \wedge
	\big( \forall t \in [0, t_f), x(t) \notin \mathcal{O} \big)
	\big\}.
\end{multline}
where \(\mathcal{X}_{\text{in}} \) is the initial state, \( \mathcal{T} \) is the target set, and \( \mathcal{O} \) is the obstacle set. This specification enforces that, starting from an initial state in the set, the system must eventually reach \( \mathcal{T} \) at some finite time \( t_f \), while strictly avoiding \( \mathcal{O} \). The goal here is to design a controller $ \mathcal{C}: \mathcal{X} \rightarrow 2^\mathcal{U}$ such that the trajectories of the closed-loop system satisfy $\Sigma^{\text{reach}}$.

\subsubsection{Symbolic Abstraction}

In ABCD, in order to enable formal reasoning and automatic controller synthesis, a \emph{finite-state abstraction} of the continuous system is constructed. This abstraction is achieved through two main steps:
i) \textbf{Discretization}: Partitioning the continuous state and input spaces into finitely many regions.
ii) \textbf{Transition relation establishment}: Establishing the relation $\mathcal{R}$ between the states of the original and abstract systems. This process results in a \emph{symbolic model} of the original system.

To discretize the continuous state space $\mathcal{X} \subseteq \mathbb{R}^n$, a uniform grid\footnote{Recent works have proposed smart or nonstandard discretization methods to enhance the efficiency of approach~\cite{calbert2024smart}. However, in this work, we focus exclusively on the classical, uniform grid-based abstraction.} is constructed using a quantization parameter $\eta \in \mathbb{R}{>0}^n$. The grid points form a lattice given by
\begin{equation}
	[\eta_x \mathbb{Z}^n] = \left\{ c \in \mathbb{R}^n \,\middle|\, \exists k \in \mathbb{Z}^n, \, \forall i \in \{1, \dots, n\}, \, c_i = k_i \eta_{x,i} \right\}.
\end{equation}
The \emph{abstract state space} $\mathcal{X}_d$ is then defined as the set of lattice points that lie within the original continuous space:
\begin{equation}
	\mathcal{X}_d = [\eta_x \mathbb{Z}^n] \cap \mathcal{X}.
\end{equation}

A \textbf{relation} \( {R}_x \subseteq \mathcal{X} \times \mathcal{X}_d \) is introduced to associate each continuous state with its corresponding abstract state. In this work, we adopt a common instantiation where the abstract state space consists of the center points of uniform, axis-aligned grid cells, and define:
\begin{equation}
	{R}_x= \{(x, x_d) \mid \|x - x_d\|_{\infty} \leq \eta_x/2, x_d \in \mathcal{X}_d\}.
\end{equation}

The same quantization procedure is applied to the input space $\mathcal{U} \subseteq \mathbb{R}^m$, resulting in a discrete input set $\mathcal{U}_d = [\eta_u \mathbb{Z}^m] \cap \mathcal{U}$ using input grid resolution $\eta_u \in \mathbb{R}_{>0}^m$ with the relation ${R}_u$

\subsubsection{Control Synthesis}

Once the finite abstraction is constructed, the system is represented as a finite transition system (FTS) \( \mathcal{S}_d = (\mathcal{X}_d, \mathcal{U}_d, \Delta) \), where
	 \( \mathcal{X}_d \) is the discrete (abstract) state space,
	 \( \mathcal{U}_d \) is the discrete input set,
	 \( \Delta \subseteq \mathcal{X}_d \times \mathcal{U}_d \times \mathcal{X}_d \) is the transition relation.

The transition relation \( \Delta \) encodes the dynamics by providing an over-approximation of the image set of every point in the abstract state-space. Specifically, for \( x_d, x_d' \in \mathcal{X}_d \) and \( u_d \in \mathcal{U}_d \), we say:
\begin{multline}
	(x_d, u_d, x_d') \in \Delta \Leftrightarrow 
	\exists x \in {R}_x^{-1}(x_d), u \in {R}_u^{-1}(u_d), t>0,\\
	: \xi_{x,u}(t) \in R_x^{-1}(x_d'),
\end{multline}
where \( \xi_{x,u}(t) \) denotes the solution to \eqref{eq:1} starting from state \( x \) under input \( u \).

Given this discrete abstraction, controller synthesis can be formulated as a reachability analysis over the transition system. The goal is to design a symbolic controller \( \mathcal{C}_d : \mathcal{X}_d \rightarrow 2^{\mathcal{U}_d} \) such that all trajectories starting from the initial set reach the target set while avoiding the obstacle set.

\subsubsection{Concretization and Implementation}

Now, the symbolic controller \( \mathcal{C}_d \) is concretized for implementation on the original system using \emph{feedback concretization}.
For any continuous state \( x \in \mathcal{X} \), the controller selects an input:
\begin{equation}
	\mathcal{C}(x) \in \mathcal{C}_d({R}_x(x)),
\end{equation}
where \( {R}_x(x) \) is the abstract state corresponding to \( x \) under the relation \( R_x \). This ensures that any input selected from \( \mathcal{C}_d \) is valid for the true dynamics of the system. As a result, the closed-loop system with controller \( \mathcal{C} \) satisfies the original specification \( \Sigma^{\text{reach}} \), under the assumption of sound abstraction.

The abstraction-based control pipeline thus enables formal synthesis of controllers that are guaranteed to meet safety and performance objectives. This framework forms the foundation of the symbolic synthesis procedures implemented in \texttt{Dionysos}.

\section{Problem Statement}

As a case study for this research, we focus on reach-avoid problems. We consider the example presented in \cite{reissig2016feedback}, which involves the dynamics of a bicycle model described by:
\begin{equation} \label{eq:bicycle_dynamics}
	f(x, (u_1, u_2)) = 
	\begin{pmatrix} 
		u_1 \cos(\alpha + x_3) \cos(\alpha)^{-1} \\
		u_1 \sin(\alpha + x_3) \cos(\alpha)^{-1} \\
		u_1 \tan(u_2)
	\end{pmatrix},
\end{equation}
where $x \in \mathcal{X} \subseteq \mathbb{R}^3$ is the state vector: the first two components represent the position of the bicycle, and the third is its orientation. The control inputs $(u_1, u_2) \in \mathcal{U} \subseteq \mathbb{R}^2$ correspond to the rear wheel velocity and the steering angle, respectively. 
The parameter $\alpha$ is the angle between the bicycle’s orientation and the velocity direction of its center of mass, computed as:
$
\alpha = \arctan\left(\tan(u_2)/2\right).
$
The inputs are constrained to the compact set $\mathcal{U} = [-1, 1] \times [-1, 1]$, while the state space $\mathcal{X}$ depends on the environment in which the bicycle operates. 

The goal of the reach-avoid problem is to synthesize a controller that drives the system from an initial set $\mathcal{X}_{\text{in}}$ to a target set $\mathcal{T}$ within finite time, while avoiding a set of obstacles $\mathcal{O} \subset \mathcal{X}$ throughout the trajectory. 

In this setting, the system dynamics are known and do not change after completing each task. Instead, it is the problem specifications (e.g., target state, and initial state) that vary from task to task. Therefore, in this work, LLMs are leveraged only at the level of problem specification extraction. Once the specifications are correctly extracted, the control problem can be solved using symbolic control techniques.

\section{LLM-Based Agents}

\begin{figure}
	\begin{center}
		\includegraphics[width=8.4cm]{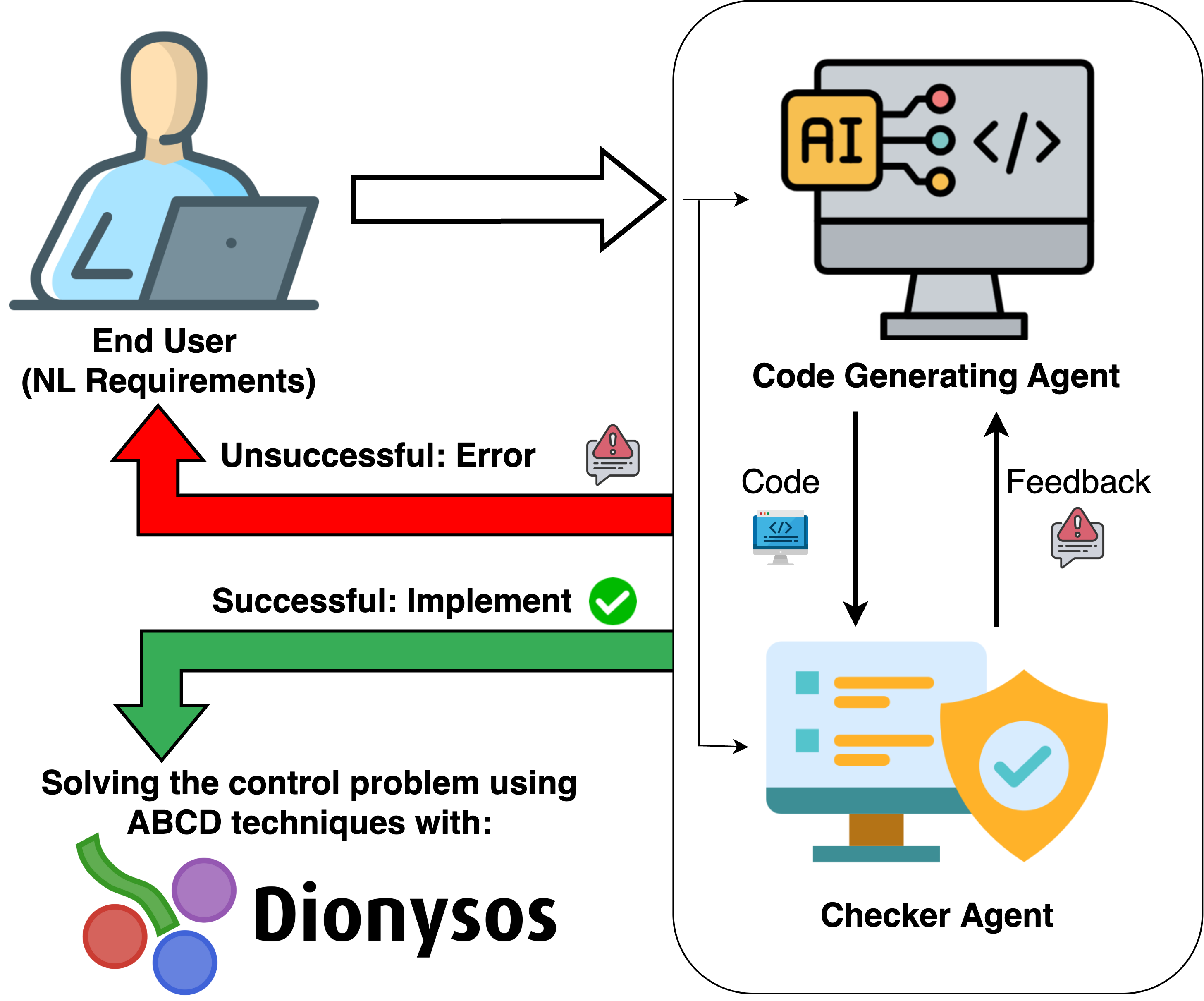}    
		\caption{Overview of the procedure with two LLM-based agents.} 
		\label{fig:flow}
	\end{center}
\end{figure}

This section introduces two LLM-based agents, both implemented using OpenAI's ChatGPT-4o in a few-shot learning setup, as illustrated in Fig.~\ref{fig:flow}. The first agent serves as a Code Generator: given a NL specification of an environment, it extracts relevant information and produces code compatible with the \texttt{Dionysos} input format for solving reach-avoid problems.

The second agent functions as a Checker. It evaluates the code generated by the first agent, identifying mismatches between the code and the original specification. It then provides targeted feedback to refine the code and improve its correctness and safety.

As illustrated in Fig.~\ref{fig:flow}, the structure of the feedback generated by the Checker Agent is identical for both internal communication (with the Code Agent) and external communication (with the end user). The difference lies in the control flow: if the Code Agent generates an incorrect implementation, the Checker Agent initiates a feedback loop with the Code Agent for a maximum of $k_{\text{max}}$ iterations (set to 2 in this work). This relatively low value was chosen based on empirical observations that, in longer loops, the Code Agent sometimes learns to bypass the checker by producing incorrect but deceptively valid-looking code. If the output remains incorrect after the second attempt, the loop terminates, and the final feedback is forwarded to the end user. Conversely, if the generated code is correct, the Checker Agent simply returns \texttt{True}, and the code proceeds directly to the implementation stage.

\subsection{Code Generator} \label{sec:4.1}

\begin{figure}
	\begin{center}
		\includegraphics[width=8.4cm]{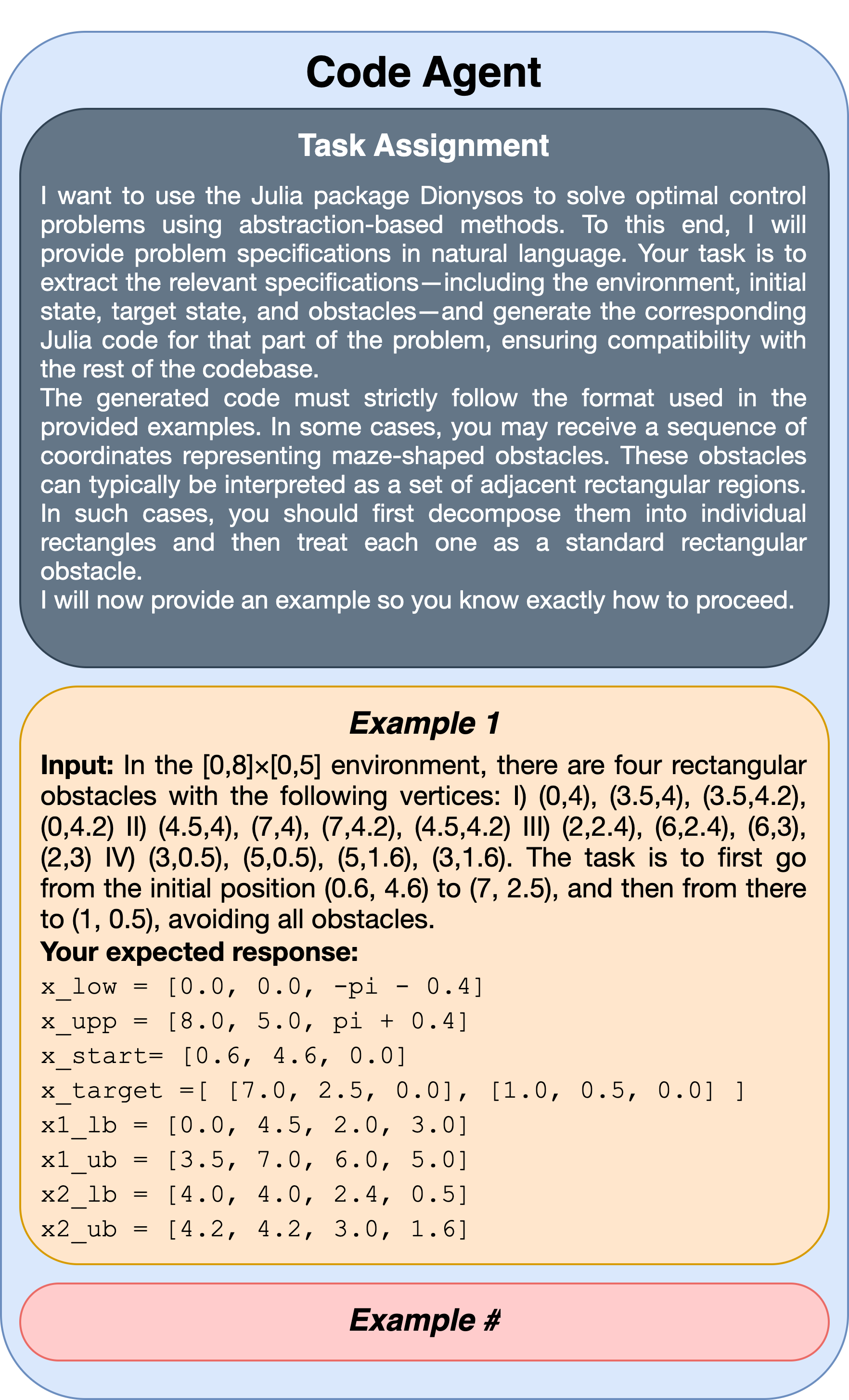}    
		\caption{Coder Agent prompt including task assignment and examples.} 
		\label{fig:Code agent}
	\end{center}
\end{figure}

This agent is designed to process NL descriptions of the operating environment, such as its boundaries, the positions of obstacles, and the requirements, such as the target(s) to be reached. The agent must extract the essential information and generate corresponding code in a specific format suitable for integration into the reach-avoid pipeline. The prompt provided to the LLM is illustrated in Fig.~\ref{fig:Code agent}

In constructing the examples for the LLM, care has been taken to include a diverse range of specifications, such as: (i) simple goals with a single target, e.g., "Go to Position\textsubscript{1} avoiding all obstacles"; (ii) sequential goals with multiple targets, e.g., "Reach Position\textsubscript{2} after visiting Position\textsubscript{1}, while avoiding all obstacles"; and (iii) safety constraints such as clearance, e.g., "You must not get closer than 1 unit to any obstacle."

Moreover, multiple NL variants are considered for defining obstacles. A rectangular obstacle may be described using: (i) the coordinates of its four vertices, (ii) the coordinates of two diagonal vertices, or (iii) its center along with the side lengths in each dimension.

The examples are constructed to reflect a representative range of phrasing and descriptive styles commonly found in NL specifications. While this variability is an inherent characteristic of NL, the goal is to provide the LLM agent with enough structured prompts to demonstrate its ability to interpret and process such variations within the scope of the task.

\subsection{Checker Agent}

The Checker Agent receives both the original NL specifications and the generated code. Using a few-shot prompting strategy with targeted examples, it is queried to identify whether the implementation aligns with the intended specifications. If the output satisfies the specification, the Checker confirms correctness and the process continues. Otherwise, it returns a structured explanation of the mismatches, prompting the Code Agent to revise its output.

\begin{figure}
	\begin{center}
		\includegraphics[width=8.4cm]{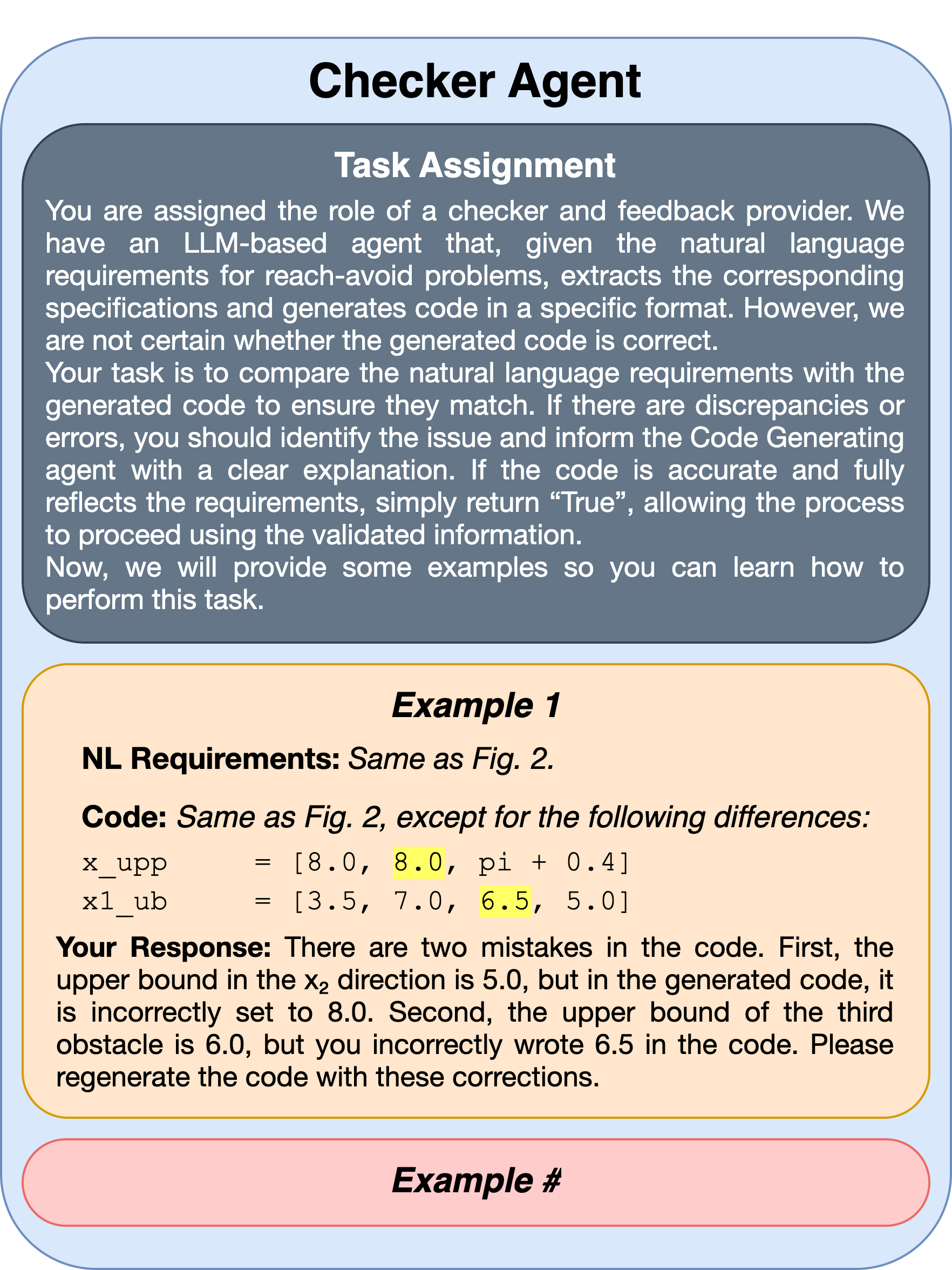}    
		\caption{Checker Agent prompt including task assignment and examples.} 
		\label{fig:CheckerAgent}
	\end{center}
\end{figure}

This process is repeated up to $k_{\text{max}}$ times if necessary. If the system fails to produce a correct version after these attempts, it stops and notifies the user. This mechanism helps reduce the chances of unsafe or incorrect behavior by preferring to halt rather than proceed with faulty code. The prompt and examples used to guide the Checker agent are shown in Fig.~\ref{fig:CheckerAgent}.

\section{Performance Evaluation}

We evaluate the methodology on a dataset of varied environments, measuring agent performance by success rate across strategies.

\subsection{Test Set}
We constructed a dataset consisting of 20 distinct environments with varying obstacle configurations. For each environment, we first manually crafted a NL specification describing the task and constraints. These specifications span a range of application scenarios, such as autonomous vehicles navigating without colliding with buildings, or mobile robots operating in warehouses while avoiding shelves.

To introduce linguistic diversity and test robustness, each original specification, as described in Sec.~\ref{sec:4.1}, was paraphrased three times, resulting in a total of 60 statements. The paraphrases vary not only in phrasing but also in how obstacles and environments are described, while preserving the original semantics. This enriches the evaluation dataset and allows us to assess the robustness of the LLM \cite{siska2024examining}: if the generated output remains consistent across semantically equivalent inputs, it indicates that the LLM handles variations in expression reliably.

\subsection{Results}

\begin{figure}
	\begin{center}
		\includegraphics[width=8.4cm]{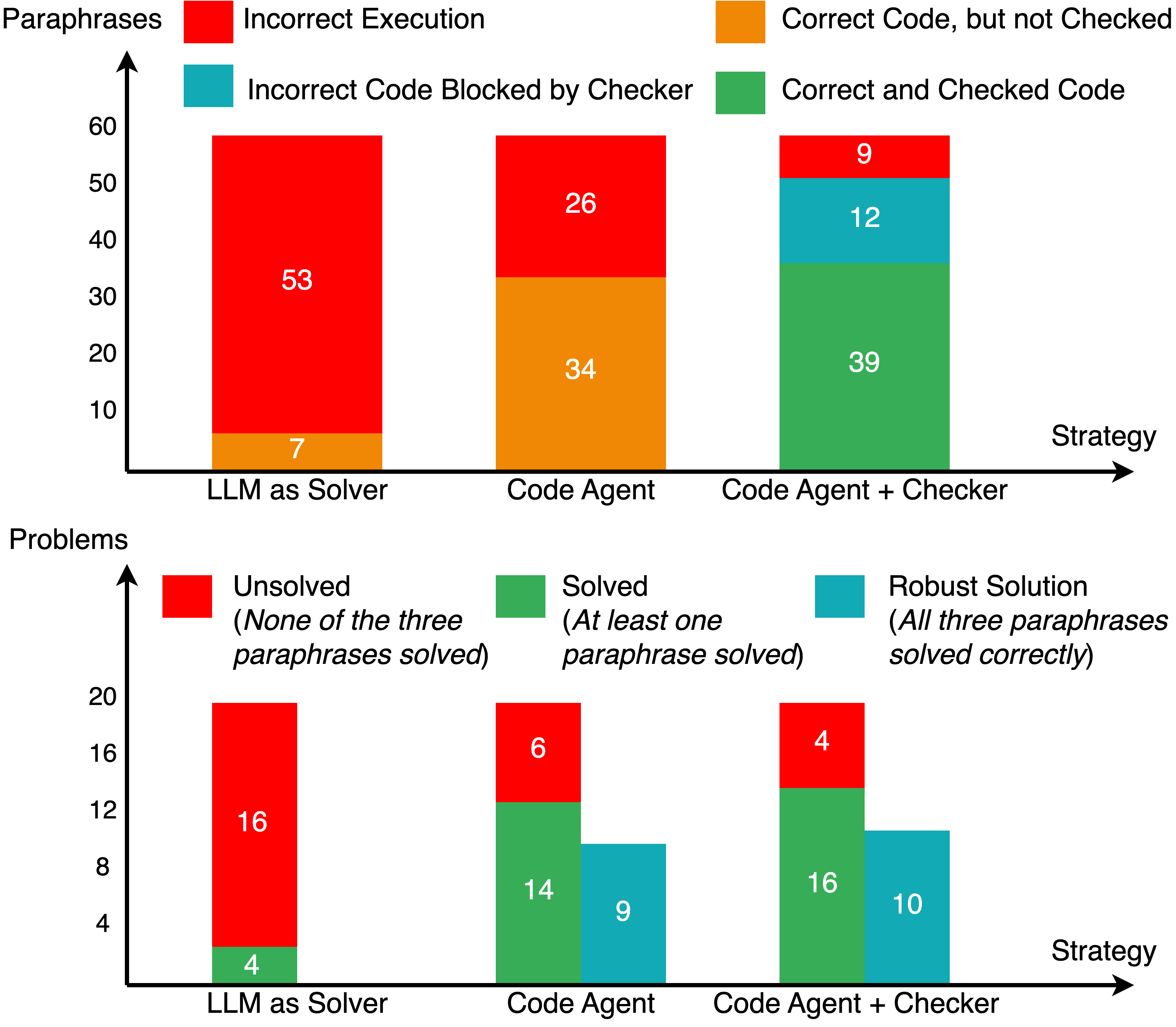}    
		\caption{Comparison of results for three strategies. Top: performance across all paraphrases (20 problems $\times$ 3 paraphrases for each). Bottom: performance over all 20 problems in the test set.} 
		\label{fig:BArChart}
	\end{center}
\end{figure}

In this section, three strategies are compared: (1) solving directly using an LLM, (2) code generation using only the Code Agent, and (3) code generation and validation using both the Code Agent and the Checker Agent (see Fig.~\ref{fig:BArChart}). This comparison highlights the effectiveness of using LLM-based agents to combine the flexibility of NL interfaces with the reliability of formal methods.

In the top chart of Fig.~\ref{fig:BArChart}, all 60 paraphrased instructions from the test set are categorized into four groups: (i) Incorrect execution: incorrect results that were executed anyway, (ii) Correct Code, but not Checked: correct results without any mechanism to verify their correctness (thus unreliable), (iii) Incorrect Code Blocked by Checker: incorrect results that were detected and blocked by the Checker Agent, and (iv) Correct and Checked Code: correct results that were verified by the Checker Agent, improving reliability.
The number of correct implementations rises significantly from 7 in the first strategy to 34 in the second and 39 in the third. While using the Code Agent alone improves performance, many incorrect codes are still executed, and none are formally verified. Incorporating the Checker Agent not only increases the number of correct solutions but also greatly improves reliability by preventing incorrect implementations, except for a few cases.

The bottom chart of Fig.~\ref{fig:BArChart} compares the number of test problems (out of 20) that were solved \emph{robustly} under each strategy. By \emph{robust}, we mean that the strategy produced a valid solution for all three paraphrased variants of a given problem. A problem is considered \emph{solved} if at least one paraphrase led to a correct code, and \emph{incorrect} if none did. Among the four problems that were solved using only the LLM (Strategy 1), none were solved robustly across all three paraphrases. In contrast, using the Code Agent alone led to 9 out of 14 problems being solved robustly (64.3\%), and the full pipeline with both agents solved 10 out of 16 problems robustly (62.5\%). It is worth noting that, while the results may vary slightly due to the non-deterministic nature of LLM outputs, the overall robustness, exceeding 60\% in both agent-assisted strategies, along with our observations, indicates that performance is relatively stable and is not expected to change significantly across repeated runs.

\begin{figure}
	\centering
	\begin{minipage}{0.43\textwidth}
		\includegraphics[width=\textwidth]{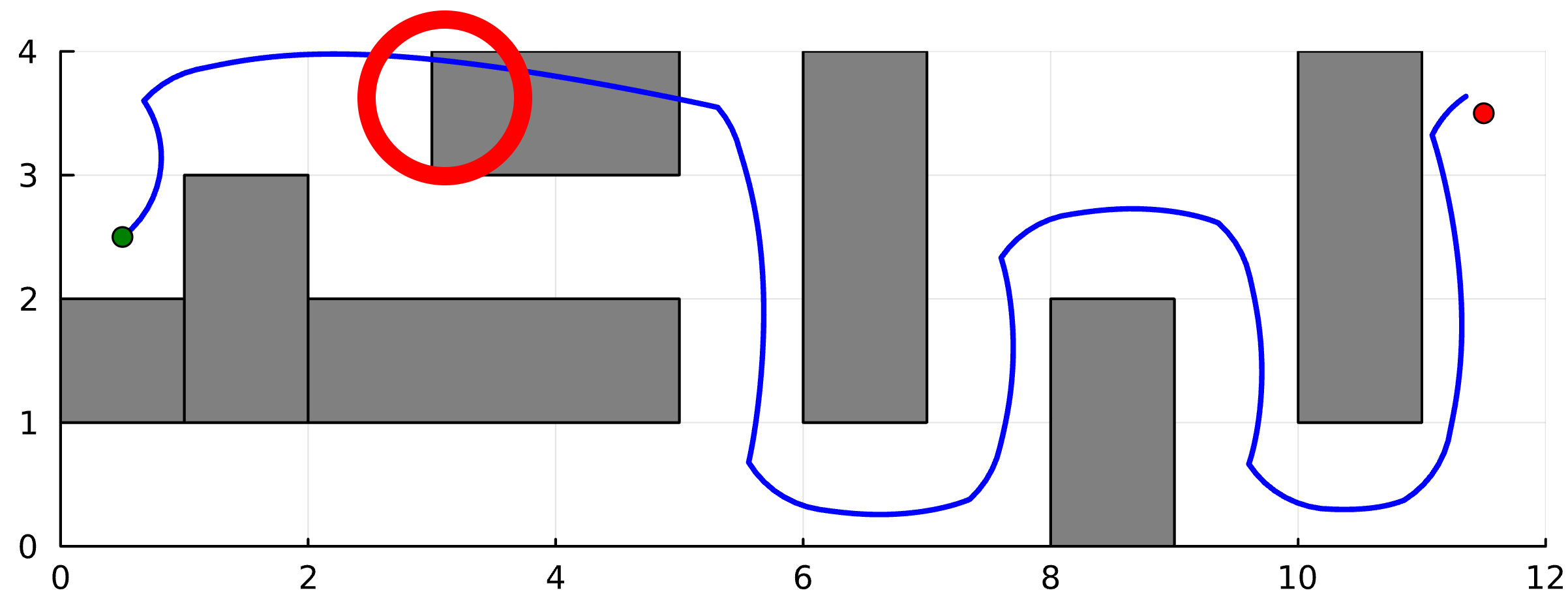}
	\end{minipage}
	\begin{minipage}{0.45\textwidth}
		\includegraphics[width=\textwidth]{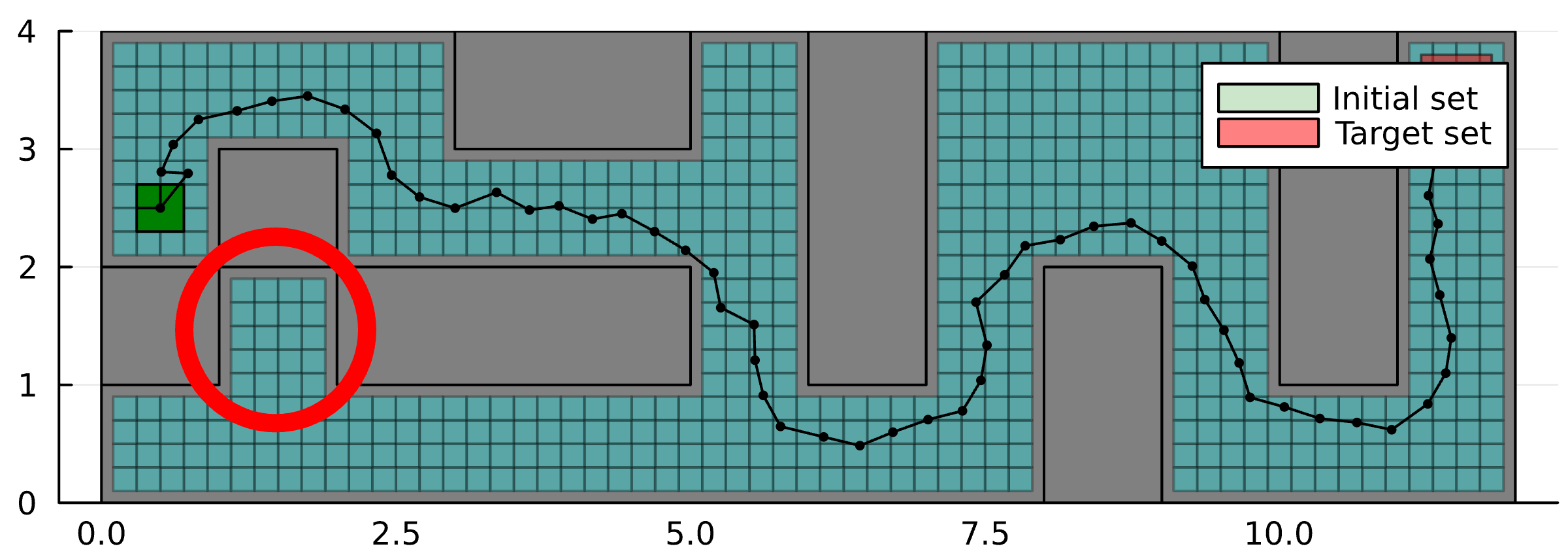}
	\end{minipage}
\begin{minipage}{0.45\textwidth}
	\includegraphics[width=\textwidth]{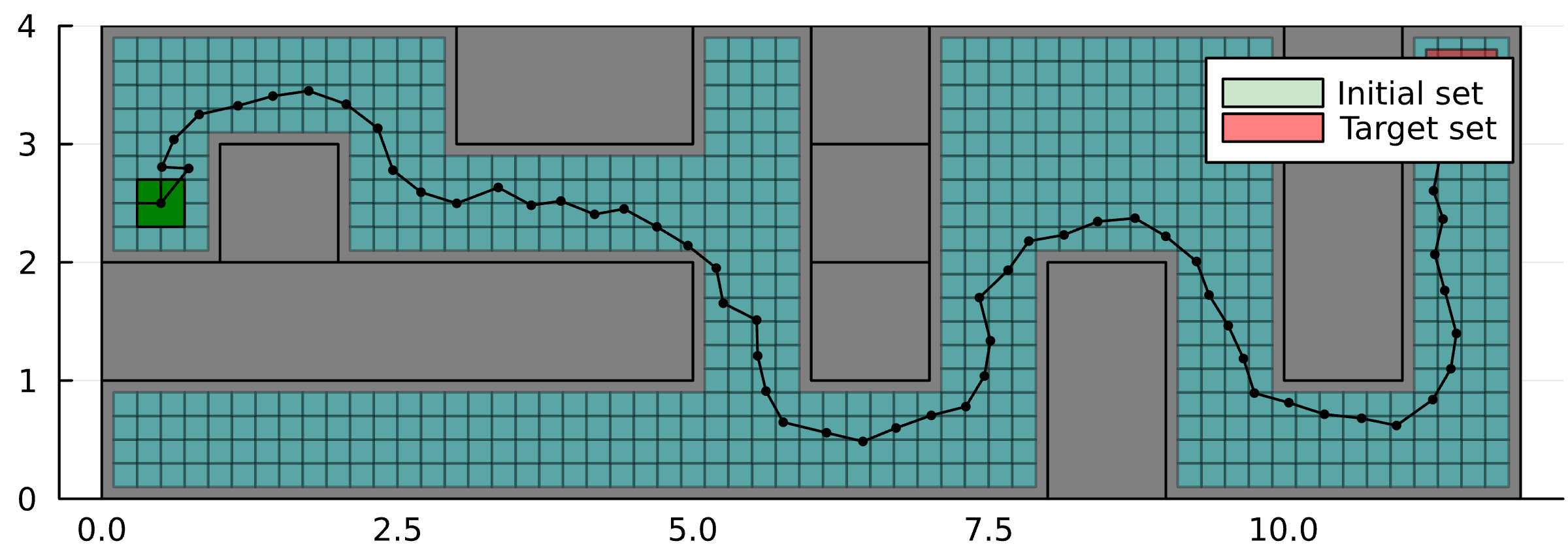}
\end{minipage}

\caption{Example trajectories from three strategies. Top: LLM used as a solver fails to avoid obstacles; Middle: Code Agent extracts incorrect specifications; Bottom: correct trajectory using symbolic control}

	\label{fig:multi}
\end{figure}

An illustrative comparison of solutions for three different cases is presented in Fig.~\ref{fig:multi}: (1) a trajectory generated by directly employing an LLM to solve the control problem; (2) an incorrect trajectory resulting from the Code Agent inaccurately representing obstacle specifications; and (3) a correct trajectory obtained using symbolic control. In case (2), the discrepancy between actual obstacles and those represented by the Code Agent is visible. Although the Checker Agent correctly identified and flagged this inconsistency, we proceeded with the flawed specification to illustrate such errors. Case (1) demonstrates that directly using an LLM to solve the control problem results in trajectories violating obstacle constraints, highlighting the limitations of using them as direct solvers.

\section{Conclusion}

This paper aimed to take a step toward a seamless, agile, and safe framework for solving safety-critical control problems using NL as the primary mode of interaction between users and the ABCD tool.. As a first step towards this goal, we have demonstrated here that LLMs are able to interface humans in the loop with a symbolic controller design tool.  An LLM-based agent was trained using a few-shot learning approach to extract reach-avoid specifications and generate the corresponding code compatible with the symbolic control platform, \texttt{Dionysos}. A second LLM agent was tasked with validating the correctness of the generated code to enhance safety and provide feedback to the first agent when necessary.

The results show that this approach significantly improves the success rate in solving reach-avoid problems. While the method lacks formal guarantees and LLM performance can be improved, it demonstrates the potential of LLMs to reduce the expertise and effort typically required for deploying formally correct control solutions.

As for future directions, the field is still in its early stages and holds significant potential. One promising direction involves fine-tuning LLMs specifically for safety-critical applications. Another promising line of work focuses on using LLMs to translate NL into formal specifications, such as temporal logic, which can then be used to solve control problems. Additionally, integrating LLMs for real-time monitoring and int eraction with system dynamics is an important direction. In many applications, requirements may evolve over time, and the system must adapt its behavior accordingly to meet the updated specifications.

\vspace{-0.5em}
\begin{ack}
\vspace{-0.5em}
Raphaël Jungers is a FNRS honorary Research Associate. This project has received funding from the European Research Council (ERC) under the European Union's Horizon 2020 research and innovation programme under grant agreement No 864017 - L2C, from the Horizon Europe programme under grant agreement No101177842 - Unimaas, and from the ARC (French Community of Belgium)- project name: SIDDARTA.
\end{ack}
\vspace{-0.3em}
\bibliography{ifacconf}             

\end{document}